\documentclass[twocolumn,pre,showpacs]{revtex4}
\usepackage{graphicx,amsfonts}

\newcommand{\norm}[1]{\left|\left| #1 \right|\right|}

\begin{document}

\title{Measurement of Indeterminacy in Packings of Perfectly Rigid Disks}
\author{Sean McNamara}
\email[Email address:]{sean@ica1.uni-stuttgart.de}
\affiliation{Institut f\"ur Computerphysik, Universit\"at Stuttgart,
70569 Stuttgart, GERMANY}

\author{Hans Herrmann}
\affiliation{Institut f\"ur Computerphysik, Universit\"at Stuttgart,
70569 Stuttgart, GERMANY}

\date{\today}

\begin{abstract}
Static packings of perfectly rigid particles with Coulomb friction
are investigated theoretically and numerically.  The problem of finding
the contact forces in such packings is formulated mathematically.
Letting the values of the contact forces define a vector in a high-dimensional
space enables us to consider the set of all possible contact forces
as a region embedded in this same space.  It is found
that the boundary of the set is connected with
the presence of sliding contacts, suggesting that a stable
packing should not have more than $2M-3N$ sliding contacts in two dimensions,
where $M$ is the number of contacts and $N$ is the number of particles.

These results are used to analyze packings generated in different
ways by either molecular dynamics or contact dynamics simulations.  
The dimension of the set of possible forces and the number of sliding
contacts agrees with the theoretical expectations.  The indeterminacy
of each component of the contact forces are found, as well as an
estimate for the diameter of the set of possible contact forces.  We
also show that contacts with high indeterminacy are located on force
chains.  The question of whether the simulation methods can represent a
packing's memory of its formation is addressed.

\end{abstract}

\maketitle

\section{Introduction}

The physics of granular materials involves two very different length
scales.  The first length scale is associated with the size of
the particles.  If we wish to give an accurate value of the density,
or describe the movement of the particles, it suffices to give the
particle positions with an accuracy of some fraction of their radii.
We call this length scale the ``kinetic'' length scale $\ell_\mathrm{kin}$,
and take it to be of order the particle radius. 
When two particles touch, inter-particle
forces at contacts are generated by tiny deformations that can be
characterized by a second length scale, that we will call 
the ``elastic'' length scale $\ell_\mathrm{el}$.

One normally has $\ell_\mathrm{el} \ll \ell_\mathrm{kin}$.  If one takes
two marbles, $\ell_\mathrm{kin}$ is about half a centimeter.
But $\ell_{el}$ is not visible to the naked eye, as one can confirm by
pushing the marbles together, and trying to observe their deformation.
Because one often has $\ell_\mathrm{el} \ll \ell_\mathrm{kin}$, it
is tempting to derive simplified numerical or theoretical approaches
by taking the limit $\ell_{el} \to 0$, corresponding to
infinitely rigid particles.  One example is the inelastic hard sphere
model, where collisions are assumed to be instantaneous.
Instead of resolving the forces during a collision, one simply
calculates the post-collisional velocities as a function of the
pre-collisional ones.

The inelastic hard sphere model opens the way to the application of
kinetic theory and the use of event driven computer simulations.
Both of these techniques have been applied successfully to a wide
variety of granular flows \cite{Stefanreview,BizonWaves,StefanEqState}.
But the neglected length scale $\ell_\mathrm{el}$
takes its revenge in an unexpected
way.  If the collisions are dissipative, ``inelastic collapse''
can occur: there can be an infinite number of collisions in finite
time \cite{collapsebernu,collapseme}.
In event driven simulations, it is necessary to
modify the simple inelastic hard sphere model to avoid this
singularity.  There are two general approaches.  In the first approach,
each particle carries a clock that records the time of its last
collision.  When two particles collide, one checks their clocks to
see if either one has had a collision less than some time $t_c$ ago.
If so, the collision dissipates no energy, otherwise, the collision
proceeds normally.  The time $t_c$ corresponds to the duration
of a collision \cite{tc}.
Alternatively, one can make the restitution coefficient
depend on impact velocity in such a way
that the energy dissipation goes to zero
as the impact velocity vanishes \cite{Bizon98}.

Another numerical method based on the approximation $\ell_\mathrm{el}\to0$
is contact dynamics (hereafter ``CD'') \cite{Fahrangthesis,Radjai96,Moreau}.  
In this method, the contact forces are calculated
by requiring them to prevent particle interpenetration and to minimize
sliding.  In this case, the neglected length scale takes its revenge
by causing indeterminacy \cite{Radjai96}.  In most cases,
there are many possible force networks that satisfy the constraints
that are imposed on them.  This raises several questions
addressed in this paper:  First of all, how big is the set of possible
contact forces?  Secondly, how are the forces chosen by CD distinguished
from all the other possible solutions?  Finally, how do the forces
chosen by CD differ from those calculated by soft-particle ``molecular
dynamics'' (hereafter ``MD'') \cite{MD}, where the particle deformations are
explicitly treated?  This paper addresses these questions.

Another recently proposed approach similar to CD is the 
``force network ensemble'' \cite{forcenet}. 
The force network ensemble is the set of all possible force networks
that could exist in a given configuration of rigid particles. 
In Ref.~\cite{forcenet}, this ensemble was sampled to obtain force
distributions, which are compared with MD simulations.  Parallels were
drawn between the force network ensemble and the ensembles of
statistical mechanics, so that one could calculate properties of
packings by averaging over the force network ensemble.  But are
all members of the ensemble equally likely to be realized?

These questions have begun to be addressed.  
For example, it has been shown that the contact forces in static
assemblies of frictionless grains can be uniquely determined \cite{Roux}.
In Ref.~\cite{Wolf}, the CD
algorithm was adapted to sample the force network ensemble,
allowing the authors to estimate its size.
The authors found that the ensemble was not uniformly sampled,
and that the force state generated by the dynamics had special
properties.  They also carried out a detailed study of the influence
of tangential friction, and showed, that indeterminacy disappears in the
limit of vanishing friction, consistent with Ref.~\cite{Roux}.

This paper takes a different, but complementary approach.
We investigate the structure of the force network ensemble mathematically,
and show (in agreement with \cite{Wolf}) that it is a convex set.
In addition, we show that the boundaries of the set are associated
with contacts where the Coulomb condition is marginally fulfilled. 
These findings place an upper limit on the
number of such contacts that can exist in a static packing, and
a lower limit on the dimension of the force network ensemble.
We show how to locate the extremal points, enabling us to calculate
the indeterminacy of the contact forces and to estimate the size
of the force ensemble network.  We also study the difference between
the MD and CD calculation methods.

This paper is organized into two main parts.  Sec.~\ref{math} presents
a mathematical formulation of
the problem of finding the contact forces in a packing of infinitely
rigid disks with Coulomb friction.  It is shown that this problem is
equivalent to finding the intersection of a cone and a linear subspace
in a high dimensional space.  
In Sec.~\ref{numeric}, we apply these ideas
numerically to static packings of about $100$ particles, and
answer several questions about the range of possible forces
that could exist in the packing, and how they are related to the
MD and CD solutions.

\section{Mathematical Formulation}
\label{math}

\subsection{Definition of the contact matrix}

We study a system of $N$ two-dimensional, circular grains
at rest under gravity $g$ in a rectangular container.
We label each grain with a unique integer $i$, $1 \le i \le N$.
Particle $i$ is characterized by its mass $m_i$, radius $r_i$, position
$\vec r_i$, velocity $\vec v_i$, momentum of inertia $I_i$, and
angular velocity $\omega_i$.  The fixed walls of the container
could be considered as particles with infinite mass, but it is
more convenient to simply leave them out of the analysis.

Let $M$ be the number of
contacts between the $N$ grains.  Each contact can also be labeled with a
unique integer $\alpha$, $1 \le \alpha \le M$.  Contact $\alpha$ is
characterized by the two touching grains $i$ and $j$.  Given the
positions of particles $i$ and $j$, it is possible to define two
unit vectors $\hat n_\alpha$ and $\hat t_\alpha$ that point in
the directions normal to and tangent to the contact, respectively.
At the contact,
the two particles exert a normal force $R_\alpha$ and a tangential 
force $T_\alpha$ on each other.

To calculate the motion of the particles, it is necessary to know
the force $\vec f_i$ and the torque $\tau_i$ on each particle
due to the contacts .  Since $\vec f_i$ and $\tau_i$ depend linearly
on the contact forces, one can write
\begin{equation}
\mathbf{f} = \mathbf{cF}.
\label{contact_matrix_forces}
\end{equation}
Here, the contact forces and the forces experienced
by the particles have been collected together into two
column vectors $\mathbf{F}$ and $\mathbf{f}$:
\begin{equation}
\mathbf{f} = \left( \begin{array}c
   f_{1x} \\ f_{1y} \\ \tau_1 \\ \vdots \\ f_{Nx} \\ f_{Ny} \\ \tau_N
\end{array} \right), \quad
\mathbf{F} = \left( \begin{array}{c}
  R_1 \\ T_1 \\ \vdots \\ R_M \\ T_M 
\end{array} \right).
\label{vectorize}
\end{equation}
Note that $\mathbf{f}\in\mathbb{R}^{3N}$ and $\mathbf{F}\in\mathbb{R}^{2M}$.
The matrix $\mathbf{c}$ has dimensions $3N \times 2M$, and is
called the contact matrix.  It contains information about which
particles touch each other, and the geometry of the contacts.
In Sec.~\ref{sec:details}, we give explicit expressions for the
components of $\mathbf{c}$.  Following \cite{Radjai96},
we call any particular value of $\mathbf{F}$
a ``contact state'' because it gives the state of all the contacts in the
granular packing.

Using this notation, we can easily write down the system of equations
that must be solved in order to find the forces in a static granular
packing under gravity.  If the particles do not move, the contact
forces must balance the gravitational acceleration:
\begin{equation}
\mathbf{f} = - \mathbf{Mg}.
\label{static1}
\end{equation}
Here, $\mathbf{M}\in\mathbb{R}^{3N}\times\mathbb{R}^{3N}$ is a diagonal
matrix containing the masses
and moments of inertia of the particles:
\begin{equation}
\mathbf{M} = \left( \begin{array}{cccccccc}
m_1 &&&&&& \\ & m_1 &&&&& \\ && I_1 &&&& \\
&&& \ddots &&& \\ &&&& m_N && \\ &&&&& m_N & \\
&&&&&& I_N  
\end{array} \right).
\end{equation}
The vector $\mathbf{g}$ contains the gravitational accelerations of all
the particles, organized in the same way as $\mathbf{f}$ in
Eq.~(\ref{vectorize}).

In contact dynamics, the unknowns are the contact forces $\mathbf{F}$,
not the forces on each particle, so we use Eq.~(\ref{contact_matrix_forces})
to re-write Eq.~(\ref{static1}) as
\begin{equation}
\mathbf{cF} = -\mathbf{Mg}.
\label{gravity}
\end{equation}
If $\mathbf{c}$ were not singular, one could find a unique solution
by inverting Eq.~(\ref{gravity}):
\begin{equation}
\mathbf{F} = -\mathbf{c}^{-1} \mathbf{Mg}.
\end{equation}
But if $\mathbf{c}$ is singular, Eq.~(\ref{gravity}) has no unique solution,
because there exist vectors $\mathbf{F}_0\ne 0$ such that
\begin{equation}
\mathbf{cF}_0=0.
\label{eigen0}
\end{equation}
Physically, this
means that there are contact states that exert
{\sl no} net force on the particles.  This is the 
source of indeterminacy in granular
packings \cite{Radjai96}.  Once we have found a solution to
Eq.~(\ref{gravity}), we can construct an infinite number
of solutions by adding multiples of $\mathbf{F}_0$.
However,  not all of these solutions are possible,
for reasons discussed in Sec.~\ref{ContactConditions}.

In general, $\mathbf{c}$ is singular, and its null space $\mathbb{C}_0$,
plays a very important role in this paper.  By considering the dimensions
of $\mathbf{c}$, one can establish a lower bound on the dimension of
$\mathbb{C}_0$.  $\mathbf{c}$ can be applied to any vector in
$\mathbb{R}^{2M}$, so its domain has dimension $2M$.  $\mathbf{c}$
maps this vector onto another vector in $\mathbb{R}^{3N}$, so its
range has dimension of at most $3N$.  Since the dimension of the
range and null space must add to the dimension of the domain, we have
\begin{equation}
\dim \mathbb{C}_0 \ge 2M-3N.
\end{equation}
If $2M\le 3N$, $\dim\mathbb{C}_0$
could vanish, but this corresponds to a coordination number of
less than $3$.  Therefore, we expect that $\dim\mathbb{C}_0 > 0$.

\subsection{Construction of the contact matrix}
\label{sec:details}

Suppose that particle $i$ and $j$ touch at contact $\alpha$.  Then
one can define a unit vector $\hat n_\alpha$
normal to the particles' surfaces at the contact:
\begin{equation}
\hat n_\alpha = \frac{\vec r_i-\vec r_j}{\left|\vec r_i-\vec r_j\right|}.
\end{equation}
To discuss tangential forces,
we must define a unit tangential vector such that
$\hat n_\alpha \cdot \hat t_\alpha=0$.  Given $\hat n_\alpha$,
there are two possibilities for $\hat t_\alpha$,
but $\hat t_\alpha$ can be uniquely defined
by imagining that the two dimensional space is
embedded in three dimensions, and writing
\begin{equation}
\hat t_\alpha = \hat n_\alpha \times \hat z,
\end{equation}
where $\hat z$ is the unit vector, pointing upwards,
perpendicular to the two-dimensional plane. 

With these definitions, it is now possible to write the forces
$\Delta \vec f_{i\alpha}$ and $\Delta \vec f_{j\alpha}$
due to contact $\alpha$ on particles $i$ and $j$:
\begin{equation}
\Delta \vec f_{i\alpha}= R_\alpha \hat n_\alpha + T_\alpha \hat t_\alpha, \quad
\Delta \vec f_{j\alpha}= -R_\alpha \hat n_\alpha - T_\alpha \hat t_\alpha,
\label{rel_vel0}
\end{equation}
But this notation 
is awkward, because it is necessary to distinguish the two
particles of the contact.  One of the particles is ``first'' (particle $i$)
and the other is ``second'' (particle $j$).  The choice of which particle is
first is arbitrary, but once the choice is made, it must not be changed.
Accordingly, we introduce the symbol $\chi_{i\alpha}$ defined by
\begin{equation}
\chi_{i\alpha} = \left\{
   \begin{array}{cl}
	1 & \mbox{if particle }i\mbox{ is first in contact }\alpha,\\
	-1 & \mbox{if particle }i\mbox{ is second in contact }\alpha,\\
	0 & \mbox{if particle }i\mbox{ does not participate in contact }\alpha.
   \end{array} \right.
\end{equation}
For each contact between two grains, one element of $\chi$ is $1$, 
and another is $-1$.  Contacts between a wall and a grain contribute
only one nonzero element to $\chi$.
Using the $\chi$ symbol, Eqs.~(\ref{rel_vel0}) 
as a single equation:
\begin{equation}
\Delta \vec f_{k\alpha}= 
\chi_{k\alpha} (R_\alpha \hat n_\alpha + T_\alpha \hat t_\alpha).
\end{equation}
This equation holds for $1 \le k \le M$, so the total force on a particle
can be written as a sum over all the contacts:
\begin{equation}
\vec f_{k}= \sum_{\alpha=1}^{M}
\chi_{k\alpha} (R_\alpha \hat n_\alpha + T_\alpha \hat t_\alpha).
\label{forceparticle}
\end{equation}
This equation can be cast in the form of a matrix multiplication.
From Eq.~(\ref{forceparticle}) and an analogous equation for the torques,
it is possible to deduce the components of
$\mathbf{c}$.  $\mathbf{c}$ is a $N \times M$ matrix of submatrices
$c_{i\alpha}$, where
\begin{equation}
c_{i\alpha} = \left( \begin{array}{cc}
\chi_{i\alpha} \hat n_{\alpha x} & \chi_{i\alpha} \hat t_{\alpha x} \\
\chi_{i\alpha} \hat n_{\alpha y} & \chi_{i\alpha} \hat t_{\alpha y} \\
0 & |\chi_{i\alpha}| r_i \end{array} \right).
\end{equation}

\subsection{The contact conditions}
\label{ContactConditions}

\begin{figure}
\begin{center}
\includegraphics[width=0.3\textwidth]{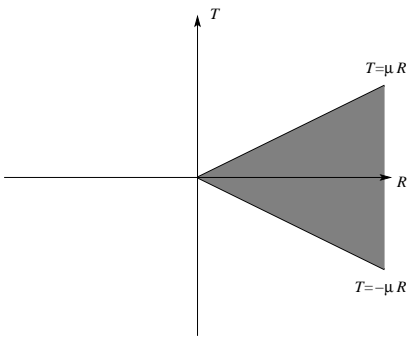}
\end{center}
\caption{\label{Kset} Cross section  of $\mathbb{K}$ cut by the $(R,T)$
of a contact.
Contacts that satisfy Eq.~(\ref{contact_conditions}) must lie in the
shaded triangular region.}
\end{figure}

Eq.~(\ref{gravity}) does not give a complete
description of motionless granular packings.  
Granular packings are nonlinear because only certain contact forces
are physically possible.  For dry granular materials
with Coulomb friction, two conditions need to be met:
\begin{equation}
R_\alpha \ge 0, \mbox { and }
\left| T_\alpha \right| \le \mu R_\alpha,
\label{contact_conditions}
\end{equation}
for $\alpha=1,\ldots M$.  The first condition says that there are no
attractive forces, only repulsive ones.  The second condition
states that the tangential force cannot exceed $\mu$ times
the normal force, where the constant $\mu$ is the Coulomb friction ratio.
Let us define $\mathbb{K}$ to be the set of all contact states
satisfying Eq.~(\ref{contact_conditions}).  In Fig.~\ref{Kset},
we show the cross section of $\mathbb{K}$, cut by the $(R,T)$ plane
of a contact.

Later in this paper, our calculations will be greatly simplified because
$\mathbb{K}$ is a convex set. 
This can be seen from Fig.~\ref{Kset}, since $\mathbb{K}$ is a cone
with its vertex at the origin.  The convexity of $\mathbb{K}$ can
also established by a simple proof.
Consider two points $\mathbf{F}_A, \mathbf{F}_B \in \mathbb{K}$. 
The points
\begin{equation}
\mathbf{F}_\lambda = \lambda \mathbf{F}_A + (1-\lambda) \mathbf{F}_B,
\quad 0 \le \lambda \le 1.
\end{equation}
lie on a straight line between $\mathbf{F}_A$ and $\mathbf{F}_B$. 
One can show that $\mathbf{F}_\lambda \in \mathbb{K}$ as well,
showing that $\mathbb{K}$ is convex.

As we shall see, contacts where an equality holds in
Eq.~(\ref{contact_conditions}), i.e.,
where $R_\alpha=0$ or $|T_\alpha|=\mu R_\alpha$, play a
special role in limiting the indeterminacy.  Contacts with
$R_\alpha=0$ are called ``non-transmitting'' contacts because
the particles touch, but exert no force on each other.
Contacts where $|T_\alpha|=\mu R_\alpha$ holds will be called ``sliding
contacts'' even if there is no relative motion.  This term is 
used because this equality holds when two particles slide relative
to one another.

\subsection{The set of possible contact states}

We are now in a position to find the contact forces
in a static granular packing and understand how indeterminacy arises.
Given an arrangement of particles and external forces acting
on each particle (e.g.~gravity), one can calculate the
contact matrix $\mathbf{c}$ and the vector $\mathbf{Mg}$.
Then one first searches for a ``particular solution''
$\mathbf{F}_1$ such that $\mathbf{cF}_1=-\mathbf{Mg}$.
These forces are necessary to cancel the external forces.
The particular solution $\mathbf{F}_1$ can be made unique
by requiring that it be orthogonal to every vector in $\mathbb{C}_0$.

In general, $\mathbf{F}_1$ will not obey the contact conditions
Eq.~(\ref{contact_conditions}), so one must find some
$\mathbf{F}_0 \in \mathbb{C}_0$ such that
\begin{equation}
\mathbf{F}_0 + \mathbf{F}_1 \in \mathbb{K}.
\label{null_space_condition}
\end{equation}
The sum $\mathbf{F}_0 + \mathbf{F}_1$ is a possible contact state.
There could be many vectors $\mathbf{F}_0\in\mathbb{C}_0$ that satisfy
Eq.~(\ref{null_space_condition}), so the solution may not be unique.
On the other hand, not all vectors from $\mathbb{C}_0$ will
satisfy Eq.~(\ref{null_space_condition}).

Let $\mathbb{F}$ be the set of all such contact states
$\mathbf{F} = \mathbf{F}_0 + \mathbf{F}_1$ satisfying
Eq.~(\ref{null_space_condition}).  Then
$\mathbb{F}$ is the set of all contact states that could be observed
in a given granular packing.  We can summarize the requirements
of a contact state by writing
\begin{eqnarray}
\mathbb{F} = \Big\{ \mathbf{F}_0 + \mathbf{F}_1\;\big|\;
    \mathbf{F}_0+\mathbf{F}_1 \in \mathbb{K},\,
    \mathbf{F}_0 \in \mathbb{C}_0,\,
    \mathbf{cF}_1 = \mathbf{Mg},\cr
	\mathbf{F}_1 \cdot \mathbf{F}_0^{(i)} = 0
		\mbox{ for } i = 1 \ldots \dim\mathbb{C}_0
\Big\}.
\label{defF}
\end{eqnarray}
Here $\{\mathbf{F}_0^{(i)},i=1\ldots\dim\mathbb{C}_0 \}$
denotes a basis of $\mathbb{C}_0$. 
In Ref.~\cite{forcenet}, $\mathbb{F}$ is called the ``force network
ensemble''.

It is clear that the set $\mathbb{F}$ is also convex.
Let $\mathbf{F}_a$ and $\mathbf{F}_b$ be members of $\mathbb{F}$.
Then the intermediate points $\mathbf{F}_\lambda$ have the form
\begin{eqnarray}
\mathbf{F}_\lambda &=& \lambda \mathbf{F}_a + (1-\lambda) \mathbf{F}_b =
   \mathbf{F}_1 + \lambda \mathbf{F}_{0,a} + (1-\lambda) \mathbf{F}_{0,b}\cr
   &=& \mathbf{F}_1 + \mathbf{F}_{0,\lambda},
\end{eqnarray}
where we have used the decompositions
$\mathbf{F}_a = \mathbf{F}_1 + \mathbf{F}_{0,a}$ and
$\mathbf{F}_b = \mathbf{F}_1 + \mathbf{F}_{0,b}$.
It can be shown that 
$\mathbf{F}_\lambda$ satisfies all the conditions in Eq.~(\ref{defF}):
$\mathbf{F}_\lambda\in\mathbb{K}$ because $\mathbb{K}$ is convex
and $\mathbf{F}_{0,\lambda} \in \mathbb{C}_0$
because $\mathbb{C}_0$ is a linear subspace.  The convexity of
$\mathbb{F}$ was first noted in Ref.~\cite{sandpiles}, and shown in
Ref.~\cite{Wolf}.

If $\mathbb{F}$ is empty, the granular packing is unstable, and the
particles will move.  If $\mathbb{F}$ contains only one point, there
is a unique contact state, and there is no indeterminacy.  Finally,
$\mathbb{F}$ can contain many points.  In this case, there is no
unique solution.

\subsection{The Boundary of $\mathbb{F}$}
\label{structure}

The boundary of $\mathbb{F}$ is analogous to the ``yield surface'' in
the elastoplasticity theory of soils \cite{Ramon}.  If a system crosses the
boundary, and leaves $\mathbb{F}$, the packing is no longer stable,
and the particles will move.  When this happens, a new contact matrix
must be constructed, including perhaps new contacts, and $\mathbb{F}$
and $\mathbb{C}_0$ will change.  For our purposes, we are interested in
the boundary of $\mathbb{F}$ because it contains the extremal points,
where the contact forces are maximized or minimized.

To investigate the structure of $\mathbb{F}$, we will use the concept
of an $n$ dimensional neighborhood of point. 
We say a point $\mathbf{F}\in\mathbb{F}$ has an $n$ dimensional neighborhood
if there exist at most $n$ linearly independent vectors $\mathbf{V}$
such that one can find $a_\mathrm{min}<0$ and $a_\mathrm{max}>0$ satisfying
\begin{equation}
\mathbf{F} + a \mathbf{V} \in \mathbb{F}, \mbox{ for all }
a_\mathrm{min} < a < a_\mathrm{max}.
\label{definterior}
\end{equation}
As an example, consider a line segment embedded in two dimensional space.
Each point on the line segment has a $1$ dimensional neighborhood, and the end
points have $0$ dimensional neighborhoods.  

Now let us apply this concept to our set $\mathbb{F}$.  Let us suppose that
that $\mathbb{F}$ contains a contact state $\mathbf{P}_0$ with no sliding or
non-transmitting contacts.  A multiple of any vector in $\mathbb{C}_0$ can
be added to $\mathbf{P}_0$, as long as it is small enough.
Therefore, there are $n = \dim\mathbb{C}_0$ linearly independent vectors
$\mathbf{V}$ satisfying Eq.~(\ref{definterior}), meaning that
$\mathbf{P}_0$ has an $n$ dimensional neighborhood.

Next suppose that we have a point $\mathbf{P}_1 \in \mathbb{F}$ with exactly
one sliding contact.  Let us label that sliding contact $\beta$ and assume
that we
have $\mu R_\beta^{(1)} = T_\beta^{(1)}$.  Points
$\mathbf{P}_a = \mathbf{P}_1 + a\mathbf{V}$ obey
\begin{equation}
\mu R_\beta^{(a)} - T_\beta^{(a)} = \mu R_\beta^{(1)} - T_\beta^{(1)} 
	+ a \left( \mu R_\beta^{(V)} - T_\beta^{(V)}\right),
\label{slidingarg}
\end{equation}
where $R_\beta^{(a)}$, $R_\beta^{(1)}$, and $R_\beta^{(V)}$ are the
appropriate components of
$\mathbf{P}_a$, $\mathbf{P}_1$, and $\mathbf{V}$ respectively.
Since contact $\beta$ is sliding in the
state $\mathbf{P}_1$, we have $\mu R_\beta^{(1)} - T_\beta^{(1)}=0$.  Now,
note that $a$ takes on both positive and negative values.  If $\mathbf{P}_a$
is to satisfy the contact conditions, we also need
$\mu R_\beta^{(V)} - T_\beta^{(V)}=0$, i.e., contact $\beta$ must be sliding
in $\mathbf{V}$ also.  Thus the sliding contact puts a constraint on the
vectors $\mathbf{V}$ that can be used.  However, given any two vectors from
$\mathbb{C}_0$, one can construct a linear combination satisfying
this constraint.  In this way,
$n-1$ linearly independent vectors can be constructed, so
$\mathbf{P}_1$ has an $n-1$ dimensional neighborhood.
Similar reasoning can be extended to show that a contact state with two
sliding contacts has an $n-2$ dimensional neighborhood, and finally,
a contact state with $n$ sliding contacts is a $0$ dimensional neighborhood,
that is, it is an extremal point.

So far, non-transmitting contacts have not been considered.  But it
is easy to incorporate them, because they
can be considered as two sliding contacts superimposed on each other,
i.e., the contact obeys $T=\mu R$ and $T=-\mu R$ at the same time.

\subsection{Quantifying indeterminacy}
\label{quantify}

The indeterminacy of the granular packing is determined by the size
and shape of the set $\mathbb{F}$.  Since $\mathbb{F}$ is a convex
set with a finite number of extremal points, one possible approach
would be to locate all its extremal points, but the large number of
such points makes this unfeasible.  Therefore, we adopt an
alternative approach.  We locate the subset $\mathbb{F}_\mathrm{ext}$
of extremal points where a component of some contact force attains
its maximum or minimum 
possible value.  Specifically, for each contact $\alpha$,
four different extremal points are found: the state where $R_\alpha$ is
maximum, then where $R_\alpha$ is minimized, and then the states
where $T_\alpha$ is maximized and then minimized.  (Note that minimizing
a normal force means making it approach $0$ as closely as possible,
but minimizing a tangential force means making it approach $-\infty$.)  

Once $\mathbb{F}_\mathrm{ext}$ has obtained, different
measures of indeterminacy can be extracted.  One possibility is to
calculated
the range of possible forces that a given contact $\alpha$ can take on:
\begin{equation}
\delta_{R,\alpha} = \frac{R_{\mathrm{max},\alpha}
	- R_{\mathrm{min},\alpha}}{\overline{m}g},\quad
\delta_{T,\alpha} = \frac{T_{\mathrm{max},\alpha}
	- T_{\mathrm{min},\alpha}}{\overline{m}g},
\label{defdeltas}
\end{equation}
where $R_{\mathrm{max},\alpha}$ and
$R_{\mathrm{min},\alpha}$ are the maximum and
minimum possible values of the normal contact force at contact $\alpha$.
To obtain a dimensionless number, we divide by
$\overline{m}g$, the average weight of a particle.  We call $\delta_R$ and
$\delta_T$ the ``local indeterminacies'' because they quantify the
ambiguity of the force at one contact.

One contact force cannot be maximized independently of the others.  One
could therefore ask how much the entire contact force state must change
when we bring one contact force from its minimum value to its maximum.
As an alternative to $\delta_R$ and $\delta_T$, one could calculate
\begin{eqnarray}
d_{R,\alpha} &=& \frac{ \norm{\mathbf{F}_{R,\mathrm{max},\alpha}
	- \mathbf{F}_{R,\mathrm{min},\alpha}}}{\overline{m}g},\cr
d_{T,\alpha} &=& \frac{ \norm{\mathbf{F}_{T,\mathrm{max},\alpha}
	- \mathbf{F}_{T,\mathrm{min},\alpha}}}{\overline{m}g}.\quad
\label{defds}
\end{eqnarray}
Here, $\mathbf{F}_{R,\mathrm{max},\alpha}$ is the contact state
where the normal force is maximized at contact $\alpha$, and the
other contact states in Eq.~(\ref{defds}) are defined analogously.
These quantities estimate the ``diameter'' of $\mathbb{F}$.
Note that $\mathbb{F}$ is embedded in a $2M$ dimensional
space, so that 2M different ``diameters'' can be calculated.
We call $d_R$ and $d_T$ the ``global indeterminacies''.

\subsection{Algorithm}
\label{algorithm}

We now give the algorithm used to
find the maximum and minimum possible forces at a given
contact.  The first step is to locate
an extremal point.  Let us suppose that we begin with
a point with an $n$-dimensional neighborhood in $\mathbb{F}$.  
Given this point, an extremal point can be
found in the following way: Pick one of the $n$ vectors from the basis
of $\mathbb{C}_0$.  Then, starting from the interior point,
move in that direction until a sliding contact is detected.  Then $n-1$
linearly independent vectors can be constructed out of the basis of
$\mathbb{C}_0$, all of which preserve the status of the sliding contact.
Pick one of these vectors, and proceed in its direction until a second
sliding contact is detected (or the first sliding contact becomes
non-transmitting).  Then $n-2$ linearly independent vectors can be built
which obey these two constraints.  Continuing in this way, we will eventually
reach a point where there are $n$ constraints, and no vectors can be
constructed that respect all of them.  This is an extremal point.

Once the extremal point has been reached, its neighboring extremal points
can each be identified.  Recall that an extremal point is characterized by
$n$ constraints arising from $n$ sliding contacts.  
If we relax one of these constraints, there is one
direction in $\mathbb{C}_0$ that respects all the other $n-1$ constraints.
If we move away from our extremal point in this direction, we will eventually
encounter a new extremal point.  This is a neighboring extremal point,
connected to the current point by an edge.  Since there are $n$ possible
constraints to relax, each extremal point will have $n$ neighbors.  Each
of these neighbors can be checked.  If none of them are ``better''
than the current point 
(in the sense that the relevant contact force is greater or less),
then the current point is the best point.  If any one
of the neighboring points is better, we move there and repeat the process.
The convex structure of $\mathbb{F}$ guarantees that there are no
local minima or maxima that would trap the algorithm.

The algorithm must deal with a number of practical difficulties.
For example, it can happen that a contact must always be sliding or
non-transmitting.  It is necessary to detect this, because such a situation
reduces the dimension of $\mathbb{F}$.  Therefore before beginning to
search for extremal points, we first try to locate a point without
sliding contacts.  This can be difficult, because the simulations often
yield points with many sliding contacts.  But given such a point,
one can construct a vector belonging to
$\mathbb{C}_0$ that preserves all the sliding contacts except one, and
then moving along this vector so that the number of sliding contacts
is reduced by one.  This process can be repeated.  Sometimes this does
not work, because the simulation yields a point in a tight, 
multi-dimensional corner.  In this
case, the contact dynamics iterative solver can be used to generate
an alternative starting point.

In about 1\% of the cases analyzed in Sec.\ref{numeric}, 
extremal states are found with a different number of
sliding contacts than expected.  This may happen when one constructs linear 
combinations that satisfy a given constraint, and by chance, satisfy
several other constraints at the same time.  Another possibility is that
it is difficult to maintain sufficient numerical accuracy during the
construction of the linear combinations.  Recall that one cannot
test strict equalities with floating-point numbers; one should always
check that equality conditions such as $T=\pm\mu R$ are satisfied
to a certain tolerance.  This may cause an occasional overestimate
of the number of sliding contacts.  However, since this situation occurs
only rarely, it does not affect the conclusions of this paper.

\subsection{Two Contacts}
\label{2Contacts}

\begin{figure}
\begin{center}
\includegraphics[width=0.3\textwidth]{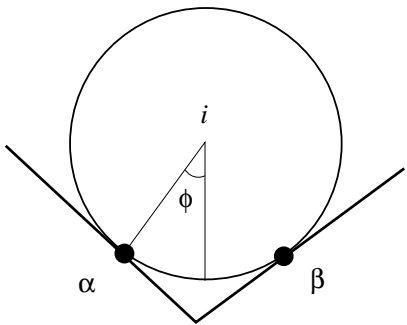}
\end{center}
\caption{\label{fig:two_contacts}The particle $i$ is supported
by two walls through contacts $\alpha$ and $\beta$.
Gravity pulls the particle downwards.  The angle $\phi$ suffices
to characterize the geometry of this simple granular packing.}
\end{figure}

As a simple application of these ideas, consider the granular
packing in Fig~\ref{fig:two_contacts}, where $M=2$ and $N=1$.
The equations of static equilibrium are
\begin{eqnarray}
R_\alpha \sin\phi + T_\alpha \cos\phi - R_\beta \sin\phi
 + T_\beta \cos\phi &=& 0,\cr
R_\alpha \cos\phi - T_\alpha \sin\phi + R_\beta \cos\phi
 + T_\beta \sin\phi &=& mg,\cr
rT_\alpha + rT_\beta &=& 0.
\label{CD_2}
\end{eqnarray}
Comparing this to Eq.~(\ref{gravity}), we have
\begin{equation}
\mathbf{c} = \left( \begin{array}{cccc}
 \sin\phi & \cos\phi & -\sin\phi & \cos\phi \\
 \cos\phi & -\sin\phi & \cos\phi & \sin\phi \\
 0 & r & 0 & r
\end{array} \right).
\end{equation}
This matrix has a null space that has at least one dimension.  Let us
find $\mathbf{F}_0$ by solving the system of equations $\mathbf{cF}_0=0$.
The result is
\begin{equation}
\mathbf{F}_0 = \left( \begin{array}c
 \sin\phi \\ \cos\phi \\ \sin\phi \\ -\cos\phi
\end{array} \right).
\end{equation}
Note that $\mathbf{F}_0$ corresponds to the horizontal components of the
contact forces canceling each other.

The particular solution can be found by solving $\mathbf{cF}_1=\mathbf{Mg}$,
and then requiring that $\mathbf{F}_0\cdot\mathbf{F}_1=0$.  The
result is
\begin{equation}
\mathbf{F}_1 = \frac{mg}{2} \left( \begin{array}c
 \cos\phi \\ -\sin\phi \\ \cos\phi \\ \sin\phi
\end{array} \right),
\end{equation}
which simply expresses the requirement that the vertical component
of the contact forces cancel gravity.
The contact states in $\mathbb{F}$ all have the form
\begin{equation}
\mathbf{F}_1 + a\mathbf{F}_0.
\label{Fset_2contacts}
\end{equation}
Applying the contact conditions puts restrictions on the values of
$a$ which are allowed.  For example, requiring $R_\alpha\ge0$ and
$R_\beta\ge0$ means that $a$ must satisfy the inequality
\begin{equation}
a \ge - \frac{mg}{2} \cot\phi.
\label{a_condition1}
\end{equation}
And the Coulomb condition becomes
\begin{equation}
\mu \frac{mg}{2} \cos\phi + \mu a \sin\phi \le
\left| - \frac{mg}{2}\sin\phi + a \cos\phi \right|.
\end{equation}
Working out the various cases connected with the absolute values,
one obtains a lower bound for $a$.
\begin{equation}
a \ge a_\mathrm{min} = - \frac{mg}{2}
   \left( \frac{\mu-\tan\phi}{1+\mu\tan\phi} \right).
\end{equation}
If this condition is fulfilled, than Eq.~(\ref{a_condition1}) is
always satisfied as well.  When $\tan\phi < 1/\mu$, there is an upper
bound for $a$:
\begin{equation}
a \le a_\mathrm{max} = \frac{mg}{2}
   \left( \frac{\mu+\tan\phi}{1-\mu\tan\phi} \right).
\end{equation}
If this condition is fulfilled, then Eq.~(\ref{a_condition1}) is
satisfied as well.
When $\tan\phi>1/\mu$, there is no upper bound on $a$; $a$ can be
arbitrarily large.

\begin{figure}
\begin{center}
\includegraphics[width=0.45\textwidth]{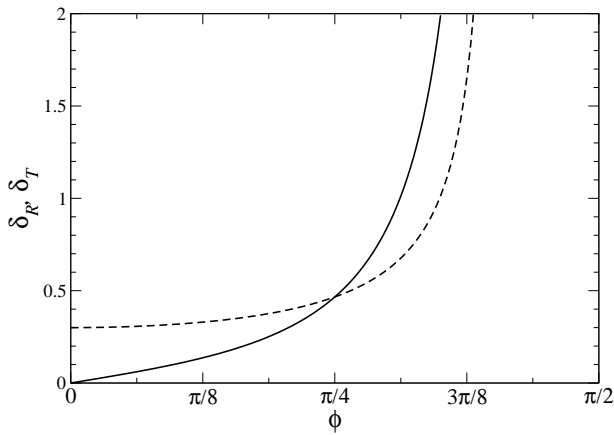}
\end{center}
\caption{\label{deltas}The measures of indeterminacy $\delta_R$
(solid line) and $\delta_T$ (dotted line).  $\phi$ is the angle shown in
Fig.~\ref{fig:two_contacts}, and $\mu$ was taken to be $0.3$.   At $\phi=0$,
$\delta_1=\mu$ and $\delta_2=2\mu$.  Both measures diverge at $\tan\phi=1/\mu$
or $\phi\approx0.41\pi$.}
\end{figure}

Now the indeterminacy of this packing can be calculated.  Selecting the
appropriate components from Eq.~(\ref{Fset_2contacts}), we have
\begin{equation}
R_\alpha = R_\beta = \frac{mg}{2}\left( \cos\phi + a \sin\phi \right).
\end{equation}
The maximum and minimum possible forces can be obtained by setting $a$ equal to
$a_\mathrm{max}$ or $a_\mathrm{min}$ respectively.  When this is done,
one obtains
\begin{eqnarray}
\delta_R &=& \frac{R_\mathrm{max}-R_\mathrm{min}}{mg} = 
   \frac{a_\mathrm{max}-a_\mathrm{min}}{mg}\sin\phi\cr &=&
   \frac{\mu\sin\phi}{\cos^2\phi-\mu^2\sin^2\phi}.
\label{Ranalytic}
\end{eqnarray}
In the same way, one has
\begin{equation}
-T_\alpha = T_\beta = \frac{mg}{2}\left( \sin\phi - a \cos\phi \right).
\end{equation}
Inserting $a=a_\mathrm{max}$ and $a=a_\mathrm{min}$, we obtain
\begin{equation}
\delta_T = \frac{\mu\cos\phi}{\cos^2\phi-\mu^2\sin^2\phi}.
\label{Tanalytic}
\end{equation}
The behavior of Eqs.(\ref{Ranalytic}) and (\ref{Tanalytic}) are shown in
Fig.~\ref{deltas}.

\section{Numerical Application}
\label{numeric}
\subsection{Overview}

\begin{figure}
\begin{center}
\includegraphics[width=0.45\textwidth]{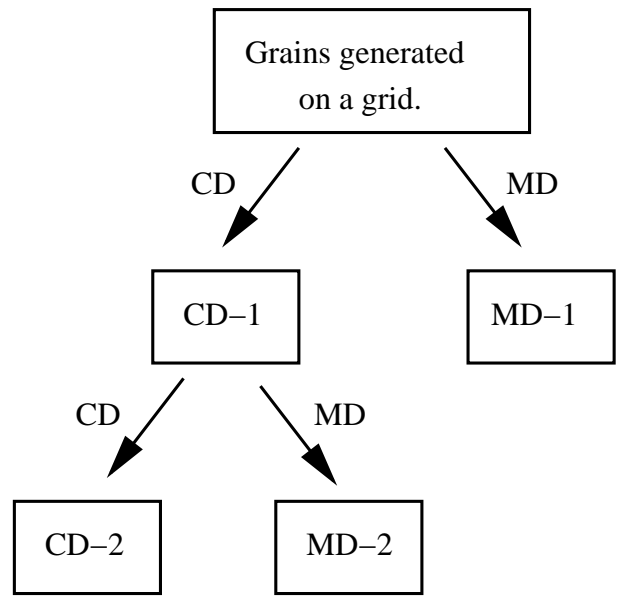}
\end{center}
\caption{ \label{simmap} A diagram showing the relationship between
the various configurations analyzed in this section.  First, $95$
grains were placed on a rectangular grid.  Then, the grains were allowed
to fall, under gravity, and settle at the bottom of the container.  This
process was simulated by both CD and MD, yielding the CD-1 and MD-1
configurations.  Then the CD-1 configuration was taken as the initial
condition for two more simulations, yielding CD-2 and MD-2.}
\end{figure}
We have investigated numerically the indeterminacy of granular packings
with $N=95$ particles.  Fig.~\ref{simmap} shows the procedure used
to generate the various configurations considered in this section.
First, $N=95$ grains were placed on a
grid inside a rectangular box of size $L_x \times L_y$.
To prevent the formation of regular arrays of particles,
the radii of the particles are uniformly distributed in the interval
$[0.7 r_\mathrm{max},r_\mathrm{max}]$. 
Then, the grains were allowed to fall, under the influence of gravity,
and form a packing at the bottom of the box.  This process was simulated
by both CD and MD, yielding the CD-1 and MD-1 configurations.  Both simulations
continue until the kinetic energy decreases to a negligible value,
or the elapsed time reaches $6\sqrt{L_y/g}$, i.e. about twice the time
a particle needs to fall a distance $L_y$.
(This second condition is needed because occasionally a particle falls to the
bottom and starts to roll with a small amount of kinetic energy that 
is enough to violate the first condition, but not large enough so that it
reaches another particle or a wall in a reasonable amount of time.)
Then the CD-1 configuration is used as the
starting point for two more simulations, yielding the CD-2 and
MD-2 configuration, according to the simulation method used.
In this second CD simulation, the initial guess for the contact forces
is $\mathbf{F}=0$.  This procedure was repeated sixty times,
yielding a set of $240$ configurations.  The $60$ initial
conditions are distinguished by choosing different particle radii
each time. 

In the MD simulations, particles interact via linear, damped springs in
both the normal and tangential directions.  When a contact becomes sliding,
it is assumed that the tangential spring remains stretched at its maximum
length.  The springs have stiffness $1.2\times10^5\;\overline{m}g/r$ and
damping constant $7.7\times 10^4\;(\overline{m}g/r)\sqrt{r/g}$.  This choice
of parameters leads to an average particle overlap of
$7\times 10^{-6}$ of a particle radius.  Usually particles are much softer
in MD simulations, but extremely hard particle were used here to approach
as closely as possible the CD simulations (average overlap:
$6\times 10^{-8}$ of a particle radius).  Due to the hardness of the particles,
the MD simulations were very slow, taking roughly $100$ times as long as the
CD ones.  In the CD simulations, the normal and tangential resitution coefficients
were set to $0$.  In all cases, the Coulomb friction ratio was $\mu=0.3$.

\begin{figure}
\begin{center}
\includegraphics[width=0.15\textwidth]{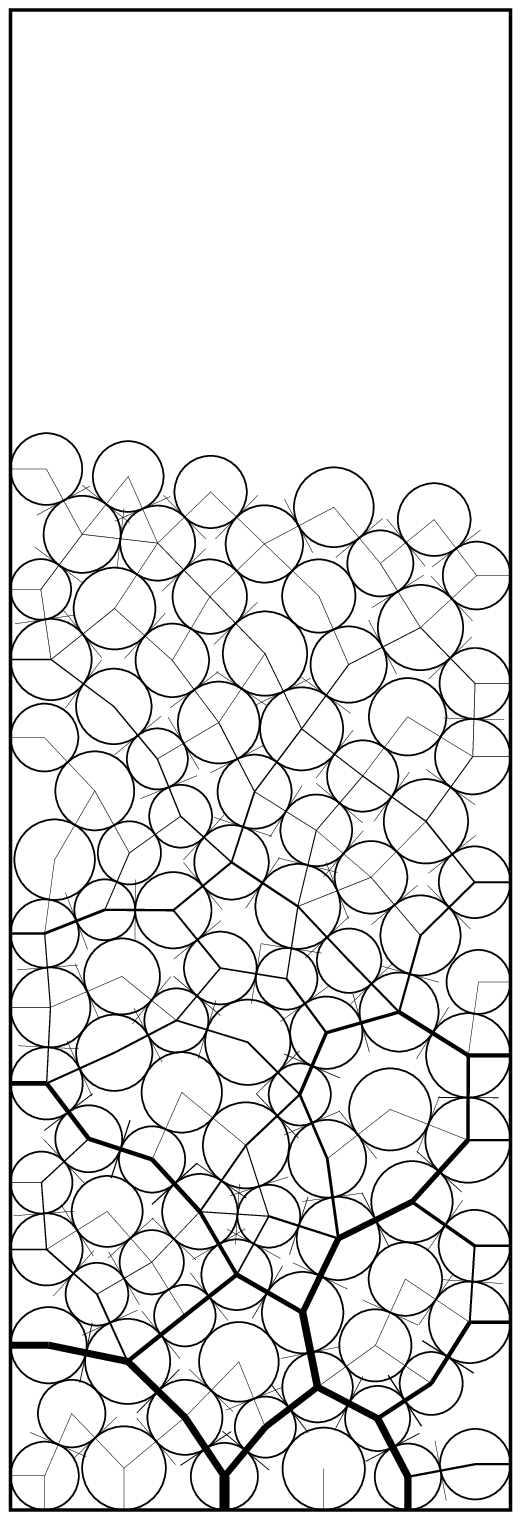}
\includegraphics[width=0.15\textwidth]{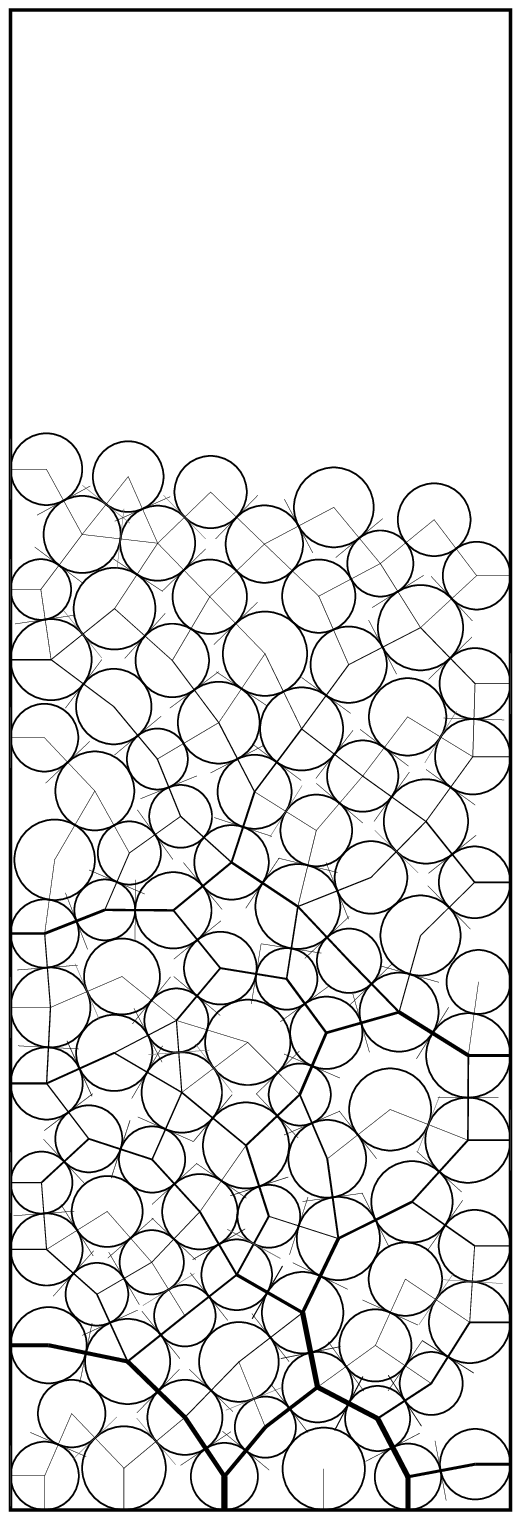}
\includegraphics[width=0.15\textwidth]{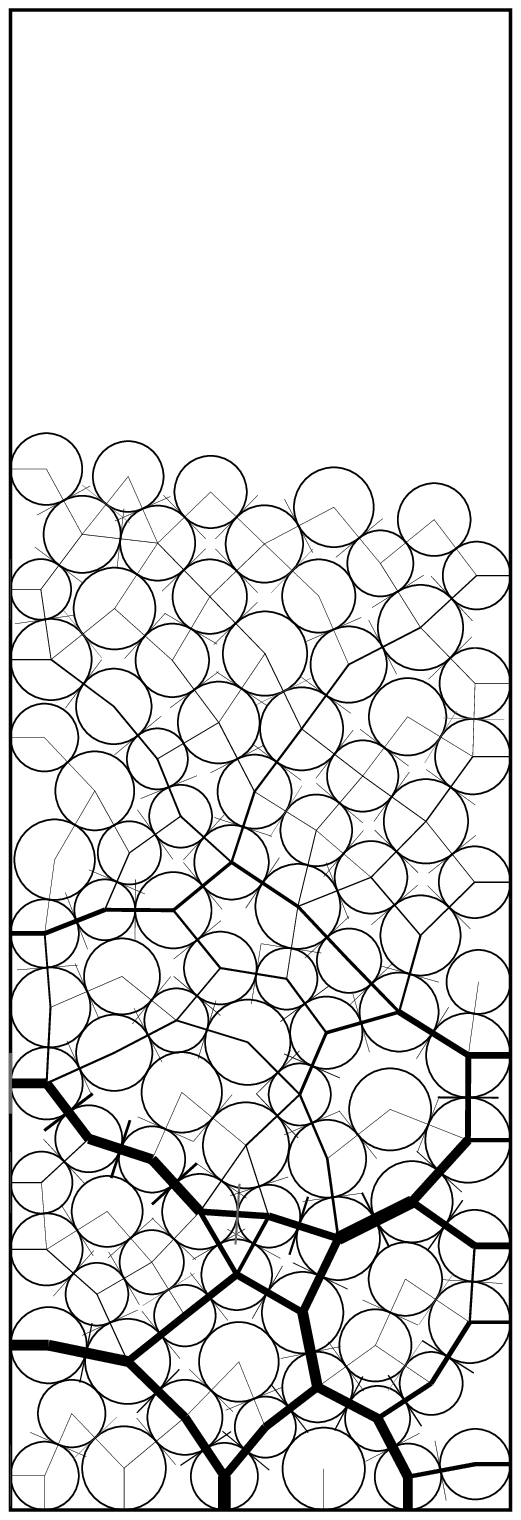}
\end{center}
\caption{\label{num_examples}
Three different contact states for a configuration with $N=95$
particles.  The CD-1 state obtained from contact dynamics
is on the left.  There are $M=160$ contacts.
Then this configuration was allowed to relax in
a molecular dynamics simulation, resulting in the MD-2 configuration
shown in the middle.
During the relaxation, the number of contacts increases to $M=166$.
The state on the right is the extremal state of the CD-1 configuration
with the maximum norm.  The thickness of the lines
connecting the centers is proportional to the normal force, and tangential
forces are shown by lines tangent to the particle surfaces.}
\end{figure}

Then the resulting configurations are analyzed.  The matrix $\mathbf{c}$
was constructed, and a singular value decomposition was used to extract
its null space.  Then the set $\mathbb{F}_\mathrm{ext}$ defined in
Sec.~\ref{quantify} is found using the algorithm presented in
Sec.~\ref{algorithm}.  Some examples of the contact states found are shown
in Fig.~\ref{num_examples}.
The left hand panel is the CD-1 state, the middle panel is the MD-2 state,
right hand panel shows one of the elements of $\mathbb{F}_\mathrm{ext}$.  
One can see already that all three
states are different, but the CD-1 and MD-2 states are closer together
than either one with the extremal state.

\begin{figure}
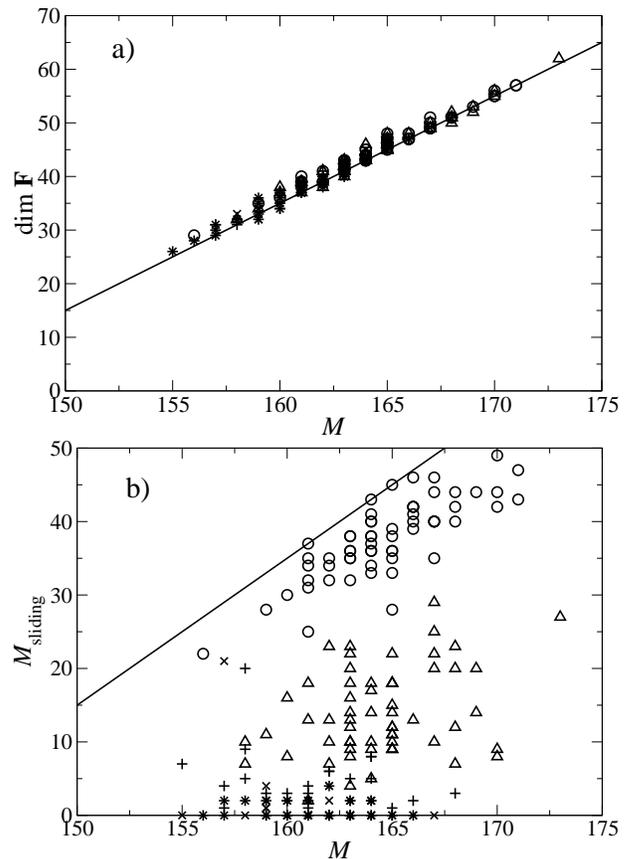

\begin{center}
\includegraphics[width=0.45\textwidth]{dimF.eps}\\
\includegraphics[width=0.45\textwidth]{Msliding.eps}
\end{center}
\caption{\label{dimF} (a) the dimension of the set of possible
contact states $\mathbb{F}$ as a function of the number of contacts $M$.
The solid line shows the lower limit $\dim\mathbb{F}=2M-3N$. 
(b) the number $M_s$ of sliding contacts as a function of $\dim\mathbb{F}$.
The straight line shows $M_s=2M-3N$.  A contact
is considered sliding if $\mu R-|T|<\epsilon \overline{m}g$,
with $\epsilon=10^{-9}$.  Non-transmitting contacts
($R,T<\epsilon\overline{m}g$) count as two sliding contacts,
consistent with our discussion in Sec.~\ref{structure}.  Four different
families of simulations are shown: triangles -- MD-1,
circles -- MD-2, crosses -- CD-1, x's -- CD-2.}
\end{figure}

Fig.~\ref{dimF}a shows the dimension of $\mathbb{F}$ as a function
of the number of contacts.  All points fall onto or just above
the line $\dim\mathbb{F}=2M-3N$, confirming the prediction
that $\dim\mathbb{F}\ge2M-3N$.  Fig.~\ref{dimF}a also shows that
$\dim\mathbb{F}$ never exceeds $2M-3N$ by more than $3$.  This means
that it is reasonable to use $\dim\mathbb{F}\approx2M-3N$ to estimate
$\dim\mathbb{F}$.  There is also no difference between the classes
of configurations, except that MD simulations tend to have more contacts
than the CD ones.  

Fig.~\ref{dimF}b shows the number of sliding contacts observed
in the different configurations.  A contact is considered sliding if
$\mu R - |T| < \epsilon \overline{m}g$.  For Fig.~\ref{dimF}b,
$\epsilon=10^{-9}$.  Non-transmitting contacts ($R,|T| < \epsilon\overline{m}g$)
are counted as two sliding contacts.
The number of sliding contacts
is always less than or equal to $2M-3N$, consistent with
our analysis of the boundary of $\mathbb{F}$ in Sec.~\ref{structure}.
The difference between the different classes of configurations becomes clear.
The CD simulations have very few sliding contacts.  The two classes
of MD simulations are also well separated from each other, with
the MD-2 configurations having the most sliding contacts.
This difference between the two configurations shows that the MD
simulation is able to retain a memory of how it was formed.  The MD-1
simulation was generated by letting the particles fall from a given
height, whereas the MD-2 simulation was started with the
particles almost in their final position.  Therefore, much more energy
was dissipated during the MD-1 simulation than the MD-2 simulations.
The difference in the number of sliding contacts is a sign of their
different histories.  

Note that the CD simulations lack this
kind of memory.  A closer examination of the CD-2 simulations reveals
that there are initially
many sliding contacts, but this number decreases rapidly to values
nearly zero, as shown in Fig.~\ref{dimF}b.
There is one exceptional simulation, indicated by the cross near
$M_s\approx20$, $\dim\mathbb{F}\approx30$, where
the system seems to be trapped in some corner of $\mathbb{F}$ and
unable to escape.  (The nearby CD-1 simulation probably represents
a similar situation, but the two points were generated by different
random number seeds, so it is probably a coincidence that they are
so close together.)

\subsection{Measurement of Indeterminacy}

\subsubsection{Local indeterminacy}

\begin{figure}
\begin{center}
\includegraphics[width=0.45\textwidth]{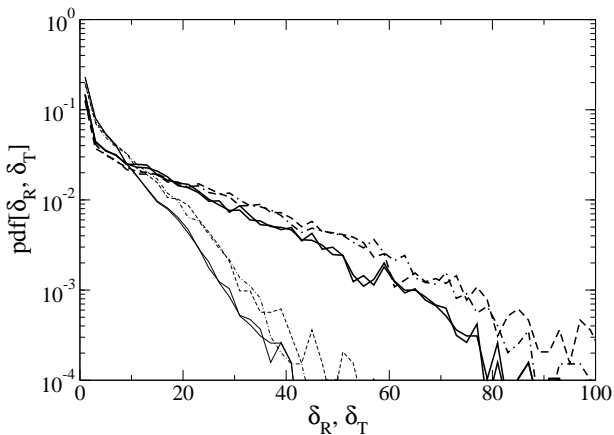}
\end{center}
\caption{\label{deltas95}Histograms of local
indeterminacy $\delta_R$ (thick lines) and $\delta_T$ (thin lines),
defined in Eq.~(\ref{defdeltas}), for the CD configurations (solid lines),
MD configurations (dashed lines), plotted semi-logarithmically.  
The thick lines show $\delta_R$ and the
thin lines $\delta_T$.  The distributions are normalized so that
their integral is unity, hence we call them probability density
functions.}
\end{figure}

The local indeterminacy presented in Sec.~\ref{quantify}
can be calculated.  In Fig.~\ref{deltas95}, we show
the distribution of $\delta_R$ and $\delta_T$ for the four
different families of configurations.  In all cases,
the distributions are exponential, except where a sharp peak
appears near $\delta_R,\delta_T=0$.  Note that similar exponential tails
are observed in contact force distributions.

The MD distributions show slightly larger indeterminacies than the
CD ones, independent of the method of generating the configuration.
No contacts were observed with infinite indeterminacies, although
it is theoretically possible, as was shown in Sec.~\ref{2Contacts}. 
Infinite indeterminacy may exist in only special packings with very small
number of particles.

\subsubsection{Global indeterminacy}

\begin{figure}
\begin{center}
\includegraphics[width=0.45\textwidth]{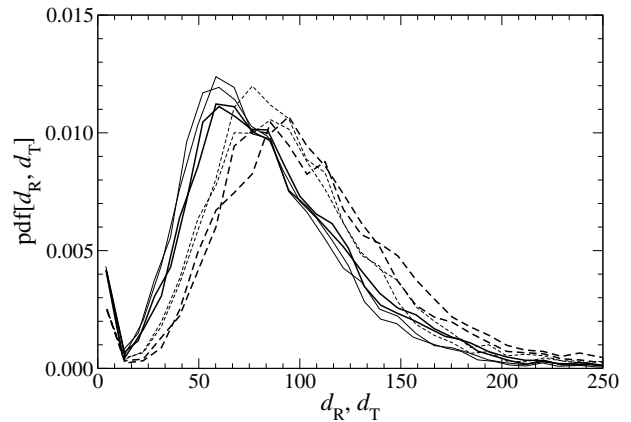}
\end{center}
\caption{\label{ds95}Histograms of global indeterminacy
$d_R$ (thick lines) and $d_T$ (thin lines),
defined in Eq.~(\ref{defds}), for the CD simulations (solid
lines), and the MD simulations (dashed lines).}
\end{figure}

Let us now consider global indeterminacy.
Histograms of the $d_R$ and $d_T$, defined
in Eq.~(\ref{defds}), are shown in Fig.~\ref{ds95} for the three different
families of configurations.
This measure of indeterminacy has very different properties from the
previous one.  In Fig.~\ref{deltas95}, the most probable values
of the indeterminacy were small, but in Fig.~\ref{ds95},
the probability density function presents two maxima,
one close to $d_R,d_T=0$, and the another well separated from the
smallest values.  The second maximum is by far the largest, so
maximizing or minimizing a contact force usually involves changing 
many forces throughout the packing, even when
the change at the contact in question is small.
The maxima in Fig.~\ref{ds95} can be taken as crude estimates of the
diameter of $\mathbb{F}$, indicating
that $\mathbb{F}$ has a diameter of approximately $60\overline{m}g$ in the CD
simulations and $75\overline{m}g$ in the MD simulations.  Consistent
with Fig.~\ref{deltas95}, the MD simulations show slightly higher
indeterminacy than the CD ones. Finally, note that
the curves for the normal and tangential components are nearly
the same in Fig.~\ref{ds95}, but clearly different in Fig.~\ref{deltas95}.

\subsubsection{Force Chains}

\begin{figure}
\begin{center}
\includegraphics[width=0.45\textwidth]{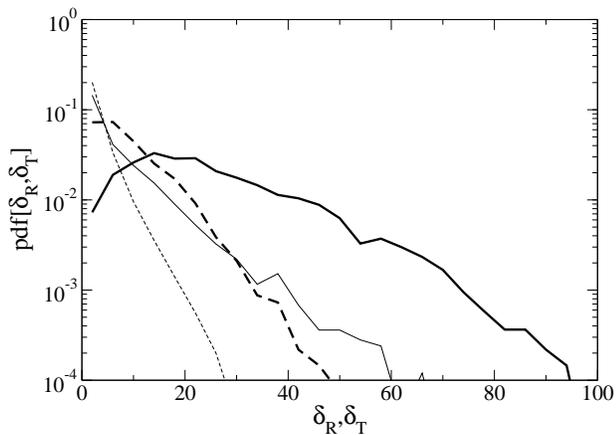}
\end{center}
\caption{\label{deltachains}Probability distributions of the
local indeterminacies $\delta_R$ (solid lines) and $\delta_T$ (dashed lines)
for contacts with $R<\overline R$
(thin lines), and for $R>\overline R$ (thick lines), where 
$\overline R=9.3\overline{m}g$ is the average normal force.  
Only data from the CD-1
simulations are shown; the others yield similar curves.}
\end{figure}

One of the most remarkable characteristics of granular packings is that
most of the force is carried by a small fraction of the contacts, which
are organized in linear structures called
force chains.  Examples of force chains are clearly visible in
Fig.~\ref{num_examples}.  This figure also suggests that indeterminacy
is also concentrated along force chains -- it is the force bearing
contacts which differ the most between the three panels. 
We can confirm this impression by dividing the contacts into two
different classes, those with above average forces and those with
below average forces.  In Ref.~\cite{ForceDist}, it was shown that
these two populations of contacts have different properties, 
the former being associated with force chains. 
(In our case, these two classes do not have exactly the same
meaning as in Ref.~\cite{ForceDist}, as the stress is not uniform throughout
the packing.  Nevertheless, it is still possible for us to isolate the
contacts in the force chains, at least in the lower part of the packing.)
In Fig.~\ref{deltachains}, we show the distributions of local
indeterminacy in the CD configurations for each class of contact.
The two classes yield quite different distributions.  The peak
at $\delta_R,\delta_T=0$ is due entirely to the
contacts with below average force, while the contacts with large
indeterminacy have above average forces.
Thus, force chains are also ``indeterminacy chains''.

\begin{figure}
\begin{center}
\includegraphics[width=0.45\textwidth]{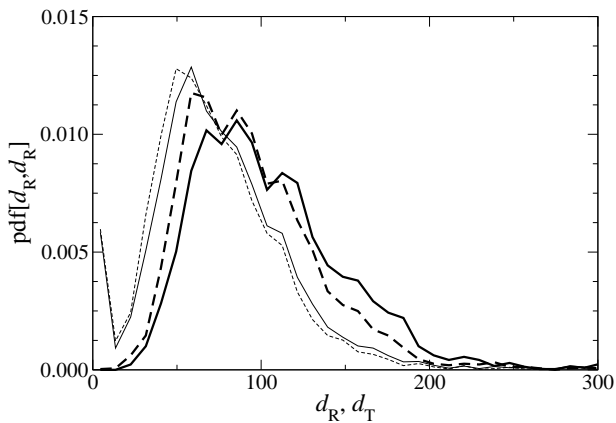}
\end{center}
\caption{\label{dchains}Probability distributions of the global
indeterminacies $d_R$ (solid lines) and $d_T$ (dashed lines)
for contacts with $R<\overline R$
(thin line), and for $R>\overline R$ (thick line), where 
$\overline R$ is the average normal force.  Only data from the CD-1
simulations are shown; the others yield similar curves.}
\end{figure}

We can examine the distribution of global indeterminacy as well.
This is done in Fig.~\ref{dchains}.
The small maximum near $d_R, d_T = 0$ is due entirely to contacts
with $R < \overline R$.  The main maximum has contributions from
both classes of contacts, but removing the weak contacts causes
the maximum to shift towards larger values.

\subsubsection{Alternative measurements of indeterminacy}

In Ref.~\cite{Wolf}, indeterminacy is measured in a different way.  
A configuration is given to the CD simulation program, and 
iterative solver of the CD simulation
is asked for a possible solution to the forces.  This solution is then
perturbed slightly, and given to the iterative solver as an initial guess,
and a new solution is obtained.  This is repeated many times, and the
force network ensemble is sampled in the same way as different
statistical mechanical ensembles are sampled in Monte Carlo simulations.
Of course, there is no guarantee that that this method will weight
appropriately the different regions of $\mathbb{F}$ or even that it
will explore all parts of $\mathbb{F}$.

We have carried out this procedure on our packings.  For each configuration,
we have obtained $500$ different solutions.  The solutions are perturbed
by multiplying all the contact forces by a random number uniformly distributed
between $0.5$ and $1.5$.  Then the center $\mathbf{F}_\circ$
of $\mathbb{F}$ can be estimated
by averaging over all $500$ points, and the radius $r$ of
$\mathbb{F}$ can be estimated by the variance:
\begin{equation}
r = \left( \frac{1}{500}\sum_{i=1}^{500}
	\left[ \mathbf{F}_i - \mathbf{F}_\circ \right]^2 \right)^{1/2}.
\end{equation}
Here, $\mathbf{F}_i$ are the solutions obtained from the iterative
solver.  Averaging over all $60$ configurations in each class, we
obtain $r=(33\pm1.4)\overline{m}g$ for the CD-1 configurations, 
$r=(34\pm1.3)\overline{m}g$ for CD-2, $r=(41\pm1.7)\overline{m}g$ for MD-1, and
$r=(40\pm1.5)\overline{m}g$ for MD-2.  These values are quite close
to half the diameter of $\mathbb{F}$ estimated from Fig.~\ref{ds95}.
In all cases, the standard deviation of $r$ is about $12\overline{m}g$,
which is also consistent with Fig.~\ref{ds95}.  This
suggests that one can indeed sample $\mathbb{F}$ in this way.

\subsection{Comparison between MD and CD}

\subsubsection{Distance between the states}

Next, we would like to examine more closely the difference between
the forces calculated by MD and CD.  One way to do this is to compare
the CD-1, CD-2, and MD-2 configurations, where the particles positions
are nearly identical.  The distance between two states $\mathbf{F}_A$ and
$\mathbf{F}_B$ is simply $\norm{ \mathbf{F}_A - \mathbf{F}_B}$.
If $\mathbf{F}_A$ and $\mathbf{F}_B$ have different
numbers of contacts, one can insert $0$'s into the appropriate
places so that they have the same dimension.

We find that the distance between CD-1 and CD-2 is $(50\pm3)\overline{m}g$,
while $(75\pm4)\overline{m}g$ separates CD-1 and MD-2.  Thus, the MD moves the
forces farther away from the initial state than CD does.  But the
MD-2 and CD-2 states are separated by only $(45\pm3)\overline{m}g$, indicating
that they move in approximately the same direction.  Note that all
these distances are less than, but of approximately the same magnitude
as, the diameter of $\mathbb{F}$.  Hence changing from CD to MD
or erasing the memory of CD causes a perturbation in the forces of
roughly the same size as their indeterminacy.

\subsubsection{Relation to extremal states}

\begin{figure}
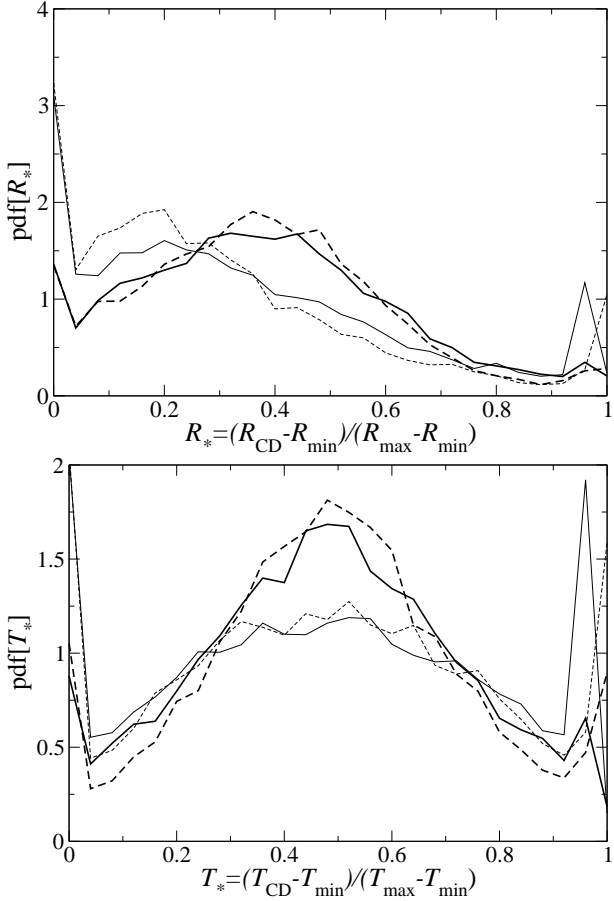

\begin{center}
\includegraphics[width=0.45\textwidth]{Rstar.eps}\\
\includegraphics[width=0.45\textwidth]{Tstar.eps}
\end{center}
\caption{ \label{RTstar} Histograms of $R_*$ and $T_*$,
defined in Eq.~(\ref{Stardef}), for the four
different configurations.  The solid lines show CD simulations, and
the dotted lines show MD simulations.  The thick lines show the
results of dropping the particles (CD-1 and MD-1), and the thin 
lines show the result of obtaining the forces with very little
particle movement (CD-2 and MD-2).}
\end{figure}

To see more precisely where the state found by the simulation
stands in relation
to the extremal states, let us consider the following quantities:
\begin{equation}
R_* = \frac{R_\mathrm{sim}-R_\mathrm{min}}{R_\mathrm{max}-R_\mathrm{min}},
\quad
T_* = \frac{T_\mathrm{sim}-T_\mathrm{min}}{T_\mathrm{max}-T_\mathrm{min}},
\label{Stardef}
\end{equation}
where $R_\mathrm{sim}$ and $T_\mathrm{sim}$ are the contact forces found
by the simulation, and $R_\mathrm{max}$ and $T_\mathrm{max}$ are the
maximum values these forces could attain in this configuration, while
$R_\mathrm{min}$ and $T_\mathrm{min}$ are the minimum forces.
$R_*=0$ means that the simulation chose the minimum possible force
while $R_*=1$ means that it chose the maximum.
In Fig.~\ref{RTstar}, we show histograms of $R_*$ and $T_*$ for the four
families of configurations.
Surprisingly, these distributions depend more on the history of the
configuration than on the simulation method.  When the particles are
dropped and allowed to settle (CD-1 and MD-1), the normal forces
are larger, relative to their minimum and maximum possible values,
than when the particles are simply placed into their final positions
(CD-2 and MD-2).  In the tangential case, the CD-1 and MD-1 states
show more values clustered around the middle of the allowed interval
than the CD-2 and MD-2 states.  On the other hand, the CD-2 and MD-2
states have more contacts at their maximum or minimum values.  These
results show that the CD method is capable of representing the history
of the packing, even though this is not reflected in the number of
sliding contacts.

\subsection{Contact Force distributions}

\subsubsection{Extremal vs Simulated states}

\begin{figure}
\begin{center}
\includegraphics[width=0.45\textwidth]{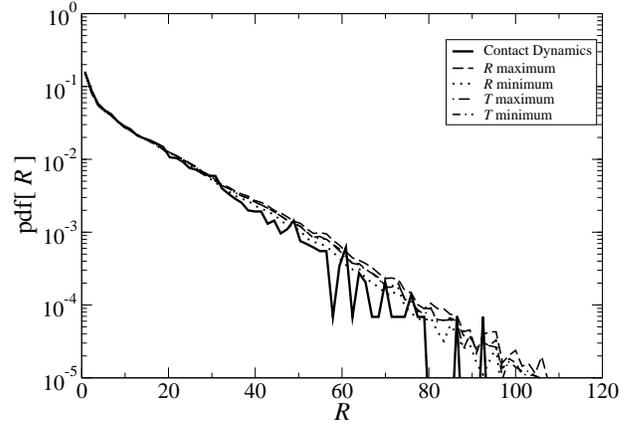}
\end{center}
\caption{ \label{RTpdf} Histogram of the normal force $R$ in the simulated
states and various classes of extremal states.  Sixty configurations
of $N=95$ particles were considered.  The curves for the extremal states
have less noise because each configuration contributes $M$ extremal
states, but only one contact dynamics state.}
\end{figure}

The contact force distributions in the various
types of contacts states are compared in Fig.~\ref{RTpdf}.
The extremal and contact dynamics distributions coincide for
$R<30\overline{m}g$, but then separate, with the extremal states
exhibiting more contacts with large forces.
This is true, even of extremal states found while minimizing $R$.
All the distributions are approximately exponential,
similar to those found in other studies \cite{ForceDist,forcenet}.
This result suggests that the exponential contact force distributions
are a property of all members of $\mathbb{F}$.  This means that
these exponential tails are probably due to some property of 
Eq.~(\ref{gravity}), and can be studied using the force network approach.

\subsubsection{MD vs CD}

\begin{figure}
\begin{center}
\includegraphics[width=0.45\textwidth]{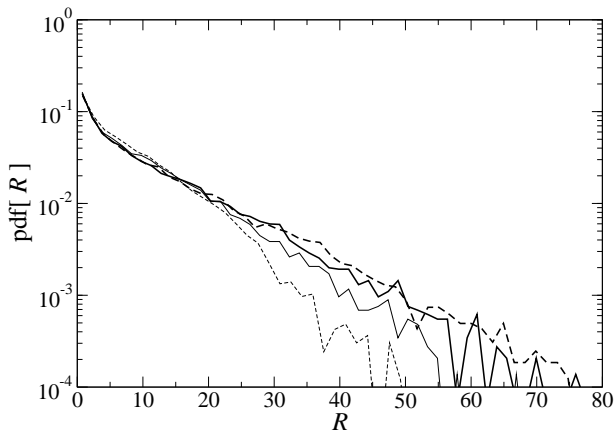}
\end{center}
\caption{ \label{RTcdmd} Histogram of the normal force $R$ for 
the different families of simulations: CD-1 -- thick solid line, 
CD-2 -- thin solid line, MD-1 -- thick dashed line,
MD-2 -- thin dashed line.}
\end{figure}
Finally, we compare in Fig.~\ref{RTcdmd} the normal contact force 
distributions for the four different families of simulations.
Both CD-1 and MD-1 yield similar curves,
obeying an exponential fall off out to the largest observed values
of $R$.  On the other hand, CD-2 and MD-2 have fewer contacts
at these large values.  This suggests that during the formation of
the packing, some of the kinetic energy is stored as elastic energy
in the force chains.  If the packing is formed with very little
kinetic energy (as in the case of CD-2 and MD-2), fewer large
contact forces are present.  

\section{Conclusions}

One conclusion that can be drawn from this work is the importance
of sliding contacts.  In Sec.~\ref{structure}, we showed that they
are associated with the boundary of $\mathbb{F}$, and hence are a sign
that the packing is close to yielding.  Furthermore, our results
suggest that there should be fewer than $2M-3N$ sliding contacts
(counting non-transmitting contacts as two sliding contacts) in a
stable packing, and no counterexamples were found among the
$240$ configurations that were examined.  If a granular
packing is slowly loaded, our work predicts that the number of sliding
contacts will increase, and reach $2M-3N$ when the packing yields.
The MD algorithm
produces many more sliding contacts than CD.  This suggests that
packings under CD are much more stable than under MD.  

We were able to calculate the local indeterminacy, that is the range
of values a given contact force can assume.  We found that the
contacts with large indeterminacy are also those contacts that make up
force chains.  Therefore, the primary origin of indeterminacy is that
the amplitudes of the force chains can change.  This result also
suggests that force chains could be understood by investigating the
null space of the contact matrix $\mathbf{c}$, since the difference
between any two allowed states belongs to this set.

The global indeterminacy measures how much the entire network
must be adjusted in order to maximize or minimize a force at a given
contact.  The reorganizations required for most contacts are significant,
even when the local indeterminacy is small.  The global indeterminacy
can also be used to estimate the diameter of $\mathbb{F}$.  This
diameter in turn allows one to appreciate the magnitude of changes
occurring within the force network.  For example, we saw that changing
the simulation method changes the force network by an amount roughly 
equivalent to the diameter of $\mathbb{F}$.  Erasing the memory of
a CD simulation changes the forces by roughly half as much.
Our estimates of the diameter of $\mathbb{F}$ are consistent with those
obtained using a Monte Carlo-like procedure to sample $\mathbb{F}$.

Finally, we made several observations about how a packing's ``memory''
is formed.  When a packing is formed violently, with much
kinetic energy, some of this energy ends up stored in the contacts.
Such packings exhibit stronger force chains and larger contact forces
than packings formed gently, with very little kinetic energy.  Both CD
and MD simulations show this effect, although it is more significant
in the MD simulations.

\begin{acknowledgments}
We thank F.~Radjai and J.J.~Moreau for help with our contact dynamics
program.  We acknowledge support from the Deutsche Forschungsgemeinschaft
Sondernforschungsbereich 382, Project C15.
\end{acknowledgments}

\bibliographystyle{prsty}

\end{document}